\documentclass[showpacs,floatfix,preprintnumbers,amsmath,amssymb,12pt]{revtex4}

\usepackage{graphicx}
\usepackage{epsfig}
\usepackage{dcolumn}% Align table columns on decimal point
\usepackage{bm}% bold math

\begin{document}

\title {Isospin fractionation : equilibrium \emph {versus} spinodal decomposition}

\author{C. Ducoin$^{(1,2)}$, Ph. Chomaz$^{(1)}$ and F. Gulminelli$^{(2)}$}

\affiliation{
(1) GANIL (DSM-CEA/IN2P3-CNRS), B.P.55027, F-14076 Caen c\'{e}dex 5, France \\
(2) LPC (IN2P3-CNRS/Ensicaen et Universit\'{e}), F-14050 Caen c\'{e}dex, France
 }

\begin{abstract}
This paper focuses on the isospin properties of the asymmetric nuclear-matter
liquid-gas phase transition analyzed in a mean-field approach, 
using Skyrme effective interactions. 
We compare two different mechanisms of phase separation for low-density matter:
equilibrium and spinodal decomposition.  
The isospin properties of the phases are deduced 
from the free-energy curvature, which contains information 
both on the average isospin content and on the system fluctuations.
Some implications on experimentally accessible isospin observables 
are presented. 
\end{abstract}


%64.60.Fr,68.35.Rh}
%64.60.Fr 	Equilibrium properties near critical points, critical exponents
%68.35.Rh 	Phase transitions and critical phenomena

%24.10.-i 		Nuclear reaction models and methods
%24.10.Pa 	Thermal and statistical models
%24.60.Ky 	Fluctuation phenomena
%25.70.Mn 	Projectile and target fragmentation
%64.70.Fx 	Liquid-vapor transitions
%65.40.Gr 	Entropy and other thermodynamical quantities

\maketitle

%%%%%%%%%%%%%%%%%%%%%%%%%%%%%%%%%%%%%%%%%%%%%%%%
%%%%%%%%%%%%%%%%%%%%%%%%%%%%%%%%%%%%%%%%%%%%%%%%
%%%%%%%%%%%%%%%%%%%%%%%%%%%%%%%%%%%%%%%%%%%%%%%%

\section{Introduction}

The nuclear interaction is a mean-range attractive, short-range repulsive force. 
Because of this, nuclear matter is expected to present a phase transition of the liquid-gas type \cite{Finn,Bertsch}. 
The nuclear-matter phase diagram has been studied in detail in the case of symmetric matter, \emph{i.e.} a system with the same number of protons and neutrons \cite{Das-PhysRep}. 
The case of asymmetric matter involves an additional degree of freedom, isospin. 
As a result, its study is more complex and allows to observe specific effects such 
as the phenomenon of isospin fractionation \cite{Xu}. 
The knowledge of these properties is of great interest for the physics of neutron-rich systems. 
Such systems have become the subject of many investigations in the recent nuclear-physics literature. 
Exotic-nucleus transport properties and multifragmentation are expected to provide constraints 
on the nuclear equation of state, in particular on the density dependence of  the symmetry energy 
\cite{Baran-PhysRep410,Tsang,Ono}. 
In astrophysics, both neutron-star structure and supernova dynamics 
are influenced by phase transitions of neutron-rich matter in a large interval of temperatures 
and densities \cite{Lattimer-PhysRep, Glendenning}. 
All these different phenomena involve excited matter at baryon densities lower than normal nuclear 
matter density : this corresponds in the phase diagram to a region of 
instability with respect to phase separation. In the case of neutron-star crusts the system lifetime
is such that one can safely modelize phase separation in terms of equilibrium. However this is 
not the case for exploding supernova cores, and even less for heavy-ion collisions. 
For such systems with a finite lifetime, the out-of-equilibrium phenomenon of spinodal
decomposition may be the leading mechanism for phase separation \cite{Chomaz-PhysRep}.

For the work presented in this paper, we have studied nuclear matter in a mean-field approach, 
using Skyrme effective interactions. Three different parameterizations are used. The more recent one, 
Sly230a \cite{Chabanat}, is used as a reference since it has been fitted to exotic nuclei 
and neutron-matter properties, which makes it particularly well-adapted 
to the description of neutron-rich matter. Two others are used as a comparison : 
SGII \cite{SGII}, which has been shown to reproduce isospin effects studied through giant dipole resonances
\cite{Catara-NPA624}, 
and SIII \cite{SIII}, which is one of the first Skyrme parametrizations 
and is not constrained for asymmetric matter. 
Contrary to Sly230a and SGII, SIII is certainly not a realistic parameterization; 
we have included it in our comparison since it 
is still used in some transport codes because of its simplicity \cite{Baran-PhysRep410}.

In section II, we present the phase diagram of nuclear matter, 
which is a representation of phase separation at equilibrium. 
Then the phase properties are considered, comparing the case of equilibrium and spinodal decomposition.
In section III, we study the direction of phase separation, 
leading to the phenomenon of isospin fractionation.
The consequences of our findings on the possible extraction of the symmetry-energy coefficient 
from isoscaling data \cite{Tsang,Ono} are explored in section IV.
In section V we discuss fluctuations in the system isotopic composition.

%%%%%%%%%%%%%%%%%%%%%%%%%%%%%%%%%%%%%%%%%%%%%%%%
%%%%%%%%%%%%%%%%%%%%%%%%%%%%%%%%%%%%%%%%%%%%%%%%
%%%%%%%%%%%%%%%%%%%%%%%%%%%%%%%%%%%%%%%%%%%%%%%%

\section{Phase diagram}

%PC 
Gibbs 
%PC 
statistical equilibrium is defined by the maximization of the system 
%PC 
Shannon
%PC 
entropy 
$S=-Tr(\hat D log \hat D)$, where $\hat D$ is the 
%PC 
many-body 
%PC 
density matrix of the system \cite{Balian}. 
Let us consider $S$ represented for a homogeneous system in the space of observables. 
 If $S$ presents a convexity in this representation, it can be maximized by phase mixing, which corresponds to a linear interpolation between two points of the space of observables. 
The coexistence region is then defined as the ensemble of points %of the observables space
 for which entropy can be enhanced by such an interpolation. 
If the system is located at a given point inside coexistence region, 
there is a unique decomposition in a couple of two other points that gives the best maximization of entropy : 
they define the two phases at equilibrium. This couple is determined by Gibbs construction \cite{Huang}, 
which consists in building the concave envelope of the entropy surface 
by linear interpolation in the space of observables : 
phases at equilibrium then belong to the same tangential plane \cite{Muller-Serot,Glendenning}. 

This geometrical condition results in the well-known equality of all intensive parameters. 
Let us consider a system occupying a volume $V$ described by a set of 
%PC observables 
observable densities %PC
$\{  a_k \}=\{ \langle \hat A_k \rangle /V \}$ %PC
%PC $\{ \langle \hat a_k \rangle \}=\{ \langle \hat A_k \rangle /V \}=$ 
controlled by the associated intensive parameters $\{ \lambda_k \}$, related to the entropy 
density %PC
$s=S/V$ by :
\begin{equation}
%PC\lambda_l=\partial_{a_l}s(\{ \hat a_k \})
\lambda_l=\partial_{a_l}s(\{ a_k \}) % PC 
\end{equation}

%Having described the system per unit volume
The pressure $P$ is related to the entropy by :
\begin{equation}
% PC \beta P= s-\sum_k \lambda_k \langle \hat a_k \rangle
\beta P= s-\sum_k \lambda_k a_k % PC 
\end{equation}
% PC where the inverse temperature $\beta$ is the intensive 
%PC parameter associated to the energy.
Let us now consider two points of the space of observables 
% PC $\{ \langle \hat a_k \rangle \}^{(1)}$ 
$\{  a_k  \}^{(1)}$ % PC 
and 
% PC $\{ \langle \hat a_k \rangle \}^{(2)}$. 
$\{   a_k  \}^{(2)}$. % PC 
Entropy has the same tangent plane at these two points if 
$\{ \lambda_k \}^{(1)}=\{ \lambda_k \}^{(2)}$ 
(equality of plane slopes) and 
$P^{(1)}=P^{(2)}$ 
% PC (equality of plane altitudes) 
 (equality of plane origins) % PC 
: these are Gibbs equilibrium conditions on the intensive parameters.

In the case of asymmetric nuclear matter, the relevant observables are neutron, proton and energy densities, 
%PC $\{ \langle \hat a \rangle \}=(\rho_n,\rho_p,E/V)$;
$\{  a_k  \}=\{\rho_n,\rho_p,e \} $; %PC
the associated intensive parameters are 
%PC $\{\lambda\}=(-\beta\mu_n,-\beta\mu_p,\beta)$
$\{ \lambda_k \}= \{ -\beta\mu_n,-\beta\mu_p,\beta \} $
where $\mu_q$ $(q=n,p)$ are the chemical potentials  %PC
and $\beta$ is the inverse temperature. %PC
To obtain the phase diagram of nuclear matter, %PC
expressions linking observables and the associated intensive %PC
parameters (i.e. equations of states)  are needed. %PC
%PC and expressions linking these quantities are needed 
%PC to obtain the phase diagram of nuclear matter.
We have worked out such relations in the mean-field approach \cite{Ducoin}. 
%and grand-canonical formalism.

%%%%%%%%%%%%%%%%%%%%%%%%%%%%%%%%%%%%%%%%%%%%%%%%%
%%%% Details sur thermo en champ moyen : cf article 1 (NPA)
%%%%%%%%%%%%%%%%%%%%%%%%%%%%%%%%%%%%%%%%%%%%%%%%%

Let us first express the average energy density of homogeneous nuclear matter, 
$e = \langle \hat{H}\rangle/V$. %PC 
%PC $\langle \hat{H}\rangle/V=
%\mathcal{H}$.
%PC E/V$. 
With a Skyrme effective interaction, 
%PC $E/V$ 
$e$ %PC 
is a functional of one-body densities only: particle densities $\rho _{q}$ and kinetic densities 
%PC $\tau _{q}=\langle \hat{p}^2\rangle_q/\hbar^2$, 
$\tau _{q}=\langle \hat{P}^2\rangle_q/\hbar^2 V$. %PC  
%PC the index $q$ representing neutrons ($q=n$) or protons ($q=p$). 
We introduce isoscalar and isovector densities :
\begin{equation}
\begin{array}{ll}
\rho =\rho _{n}+\rho _{p}\;, & \;\tau =\tau _{n}+\tau _{p} \label{EQ:rho} \\
\rho _{3}=\rho _{n}-\rho _{p}\;, & \;\tau _{3}=\tau _{n}-\tau _{p}
\end{array}
\end{equation}
In the case of homogeneous, spin-saturated matter with no
Coulomb interaction, the energy density can be written as :
\begin{equation}
%PC \frac{E}{V}
e
                    =\frac{\hbar ^{2}}{2m}\tau 
			+C_{0}\rho ^{2}+D_{0}\rho _{3}^{2}
			+C_{3}\rho ^{\sigma +2}+D_{3}\rho ^{\sigma }\rho _{3}^{2}
			+C_{eff}\rho \tau +D_{eff}\rho _{3}\tau _{3}
\end{equation}
where coefficients $C_i$ an $D_i$ are linear combinations 
of the standard Skyrme parameters \cite{Chabanat,Ducoin}
%PC >
\[
\begin{array}{ll}
C_{0}&= \  \ 3t_{0}/8 \\  
D_{0}&=- t_{0}(2x_{0}+1)/8 \\ 
C_{3}&= \  \ t_{3}/16 \\ 
D_{3}&=-t_{3}(2x_{3}+1)/48 \\ 
C_{eff}&= \ \  [3t_{1}+t_{2}(4x_{2}+5)]/16  \\ 
 D_{eff}&= \  \ [t_{2}(2x_{2}+1)-t_{1}(2x_{1}+1)]/16
\end{array}
\]
.
% < PC 

Mean-field approaches describe nucleons as independent particles. Each particle of type $q$ is submitted to the mean field $\hat{W}_{q}$ defined by the relation 
$\delta \langle \hat{H}\rangle =Tr(\hat{W}_{q}\delta \hat{\rho_{q}})$,
%PC >
 where $\delta \hat{\rho_{q}}$ represents any variation of the one-body density matrix associated with the particle $q$, 
 % < PC
 leading to the expression :

\begin{equation}
\hat{W}_{q}=\frac{\partial 
%PC \epsilon
e %PC 
}{\partial \tau _{q}}
\frac{\hat{p}^{2}}{\hbar ^{2}}+\frac{\partial 
%PC \epsilon
e %PC 
}{\partial \rho _{q}} 
=\frac{1}{2m_{q}^{*}}\hat{p}^{2}+U_{q} 
\label{EQ:MeanField}
\end{equation}

The mean field is composed of a potential term 
$U_{q}=\partial _{\rho _{q}} 
%PC \epsilon
e %PC 
$, 
and a kinetic term involving an effective mass $m_{q}^{*}$ 
defined by $\hbar^{2}/2m_{q}^{*}= \partial_{\tau _{q}} 
%PC \epsilon
e %PC 
$.

This determines individual energy levels $\epsilon_{q}^{i}=\frac{p_{i}^{2}}{2m_{q}^{*}}+U_{q}$
%PC >
with the usual quantification conditions on the momenta provided by the boundary conditions.
% < PC . 
In the grand-canonical approach, for a given temperature $T=\beta^{-1}$, 
these levels are occupied according to Fermi-Dirac statistics :
\begin{equation}
n_{q}(p)=1/ \left[ 1+exp(\beta (p^{2}/2m_{q}^{*}+U_q-\mu _{q}) \right]
\label{EQ:distribution}
\end{equation}
where $\mu_q$ is the chemical potential of the particles of type $q$
%, \emph{i.e.} the intensive parameter associated to the density $\rho_q$
.

All one-body densities are then given by Fermi integrals. In particular 
the 
%PC 
independent-particle 
%PC mean-field grand-canonical 
partition sum can be expressed 
in the same way as for an ideal gas:
%
%\begin{equation}
%\frac{\ln Z_{0}^{q}}{V}=2 \int_{0}^{\infty }
%\ln(1+e^{-\beta (\frac{p^{2}}{2m_{q}^{*}}-\mu _{q}^{\prime })})
%\frac{4\pi p^{2}}{h^{3}} dp
%=\frac{\hbar^2}{3m_{q}^{*}}\beta \tau _{q}
%\end{equation}
%
\begin{equation}
\frac{\ln Z_{0}}{V}=2 \sum_q \int_{0}^{\infty }
\ln(1+e^{-\beta (\frac{p^{2}}{2m_{q}^{*}}-\mu _{q}^{\prime })})
\frac{4\pi p^{2}}{h^{3}} dp
=\sum_q \frac{\hbar^2}{3m_{q}^{*}}\beta \tau _{q}
\end{equation}
where $\mu _{q}^{\prime }=\mu_q-U_q$.
%PC > 
At this point it is important to recall that the independent particle partition sum
$Z_0$ should not be confused with the mean-field grand-canonical 
partition sum  which is related to the mean-field pressure $\log Z=\beta P V$ and can be computed from the mean-field entropy using $\log Z= S - \beta (<\hat H> - \mu_p <\hat N_p> - \mu_n <\hat N_n>) $. 
% < PC     

For fixed values of $\rho_n$ and $\rho_p$ at a given temperature, chemical potentials 
%PC as well as kinetic densities 
are obtained
by iteratively solving the self-consistent relations between $\rho_q$, $\mu_q$ and $m_{q}^{*}$. 
We can then 
%PC >
compute the kinetic density and 
% < PC
calculate the entropy as a function of densities. 
Indeed, the mean-field 
%PC > 
solution comes from a variational approximation to the entropy  so the mean-field entropy $S_{0}$ provides the best independent-particle approximation to the exact entropy.
% < PC  
%PC approximates the real entropy $S$ with the corresponding 
%PC mean-field entropy $S^{0}$ 
%PC >
The mean-field entropy can easily be derived from the independent-particle partition sum and the average of the mean-field potential $\langle\hat{W}\rangle_{0}$ so we get  
% < PC
\cite{Vautherin} :
\begin{equation} 
S/V \simeq S_{0}/V=\ln Z_{0}/V+\beta(\langle\hat{W}\rangle_{0}/V-\mu_{n}\rho_n-\mu_{p}\rho_p) 
\end{equation}
which gives for a fixed $\beta$ the entropy $s=S/V$ as a function of neutron and proton densities.

%%%%%%%%%%%%%%%%%%%%%%%%%%%%%%%%%%%%%%%%%%%%%%%%%
As we have already mentioned, the construction of the concave envelope of $s(e,\rho_n,\rho_p)$ 
implies Gibbs conditions for each couple of phases at equilibrium : $\beta^{(1)}=\beta^{(2)}$, $\mu_n^{(1)}=\mu_n^{(2)}$, $\mu_p^{(1)}=\mu_p^{(2)}$ and $P^{(1)}=P^{(2)}$. 
Let us now introduce the constrained entropy defined by
\begin{equation} 
s_c=-\beta f = s-\beta  %PC E /V. 
e. %PC 
\label{s_c}
\end{equation}
%PC 
which
%
%PC It 
corresponds to the opposite of the free energy $f$, divided by the temperature. 
Gibbs construction can then be performed by constructing the concave 
envelope of $s_c$ in $(\rho_n,\rho_p)$ plane at a fixed temperature $T=\beta^{-1}$.
%We can represent it in $(\rho_n,\rho_p)$ plane at a fixed temperature $T=\beta^{-1}$ : %equilibrium conditions are then reproduced by constructing the concave envelope of the %resulting surface.
Indeed, $\beta$ is imposed to be constant, and the definition of $s_c$ ensures that its tangent-plane slopes and altitude give respectively the chemical potentials and the pressure :
\begin{eqnarray}
\left( \partial_{\rho_q} s_c(\rho_n,\rho_p) \right)_{\beta}=\partial_{\rho_q} s(e,\rho_n,\rho_p)=-\beta \mu_q\\
s_c+\beta(\mu_n\rho_n+\mu_p\rho_p)=s-\beta(e-\mu_n\rho_n-\mu_p\rho_p)=\beta P
\end{eqnarray} 
%
% figure 1
%
\begin{figure}[htbp]
\includegraphics*[width=1\linewidth]{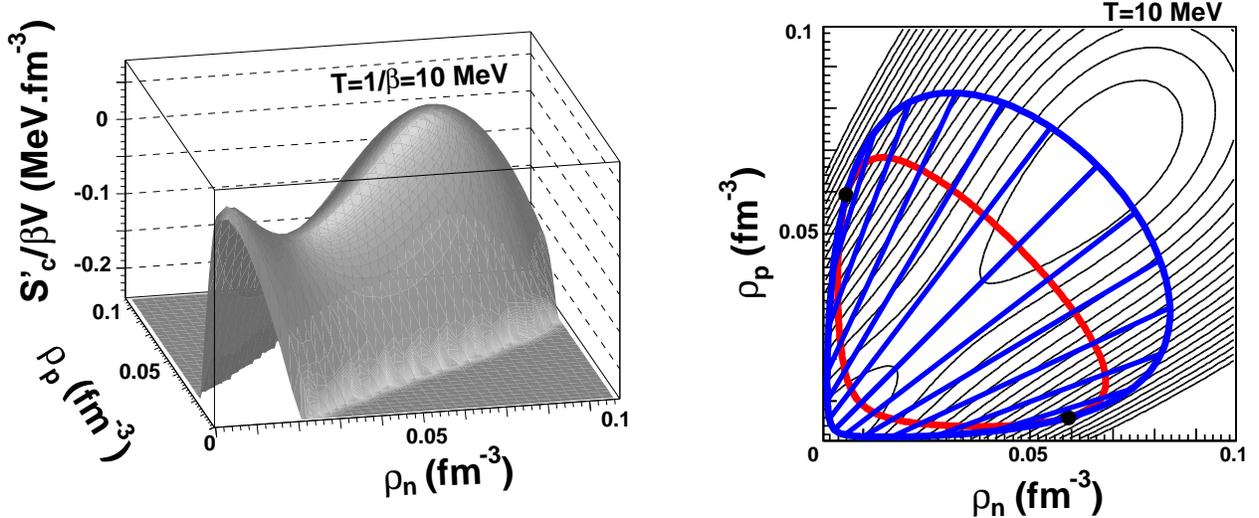}
\caption
	{
	Constrained entropy at $T=10 MeV$, corrected by
	%PC a slope 
	$\beta \mu_s\rho$ %PC 
	in order to emphasize curvature properties.
	%PC >
	It is shown on the left in perspective, and on the right as a contour plot. 
	On the right,
	% < PC Right side: projection on the proton and neutron density 
	%PCplane. The entropy surface is represented by level lines. 
	coexistence and spinodal boundaries are added as thick lines. 
	Outer curve: coexistence. Straight lines link selected couples 
	of phases at equilibrium.
	Inner curve: spinodal. Dots: critical points.
	}
\label{cs-bidim}
\end{figure}
The introduction of the constrained entropy permits then 
to reduce the dimensionality of the problem
%PC >  
since we can work at constant temperature.
% < PC  
The constrained entropy is represented in fig.\ref{cs-bidim}, 
at a temperature of $10$ MeV. 
In this figure, a slope has been subtracted in order 
to emphasize curvature properties: the surface represented is 
$s'_c=s_c  + \beta \mu_s\rho$, 
%PC  $s'_c(\rho_n,\rho_p)=-f/\beta + \mu_s\rho$,
where %PC  $\rho=\rho_n+\rho_p$,  and 
$\mu_s$ is the chemical potential of symmetric matter at the transition.
The constrained entropy is symmetric with respect to the axis $\rho_n=\rho_p$,
reflecting the invariance of nuclear interaction with respect to neutron-proton exchange.  

We observe that the surface obtained presents a region of positive convexity : this is the case if we fix any temperature under the critical temperature of symmetric matter, which is, in the Sly230a parameterization, $T_c=14.54 MeV$ (for comparison, we get $T_c=14.46 MeV$ with SGII and $17.96 MeV$ with SIII). The concave-envelope construction performed on this surface links couples of phases at equilibrium. The ensemble of these points forms the two branches of the coexistence curve, one at lower density (gas) and the other at higher density (liquid). They join at two critical points, where the difference between the two phases disappears. 

The coexistence curve is the boundary of the coexistence region, inside which a system at equilibrium is decomposed into two phases situated on each branch. The region presenting a positive convexity is the spinodal region.
%PC > 
In this region, infinitesimal density fluctuations may lead to an increase of entropy. The system is locally unstable. The spinodal region  
% < PC , which 
is by construction inside coexistence. Coexistence and spinodal boundaries are tangent at the critical points. These curves are represented on the right part of fig.\ref{cs-bidim}.
%
% figure 2
%
\begin{figure}[htbp]
\includegraphics*[width=1\linewidth]{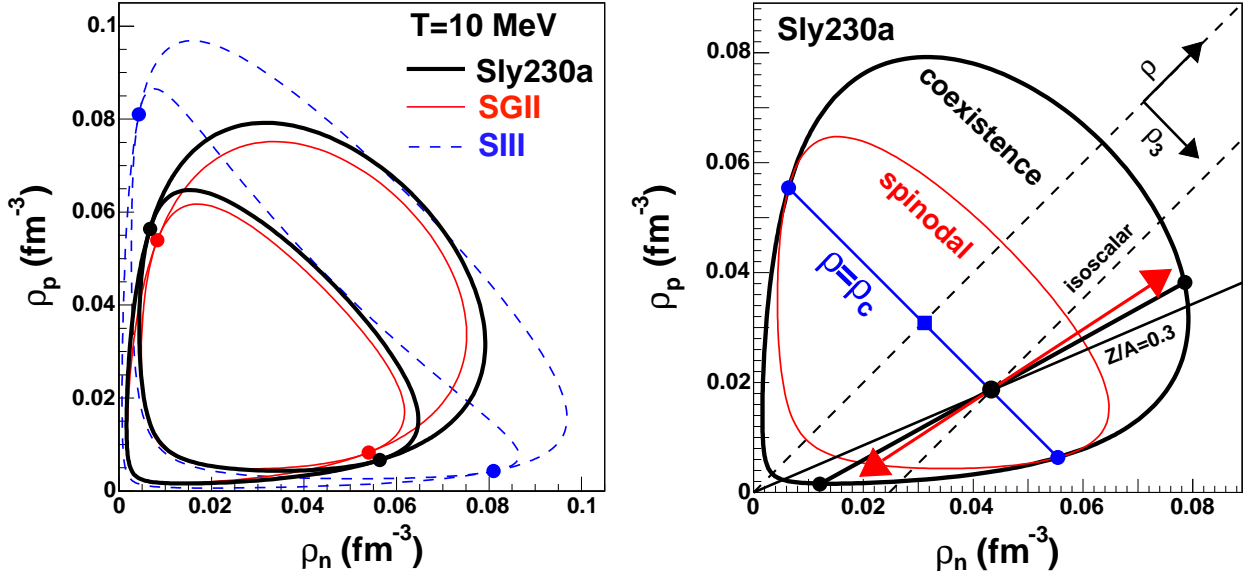}
\caption
	{
	Coexistence and spinodal curve in the density plane at $T=10 MeV$.
	Left: comparison between 
	different %PC 
	Skyrme parameterizations : Sly230a (thick lines), 
	SGII (thin lines) and SIII (dashed lines).
	Right: illustration of phase separation for a system of proton fraction 
	$Z/A=0.3$ inside instability region (\emph{cf} section III). 
	} 
\label{cs-plan}
\end{figure}
The phase diagram of the three different Skyrme forces 
is compared for the same temperature $T=10$ MeV on the left part of fig.\ref{cs-plan}.
Sly230a and SGII give close results, 
while SIII shows an atypical behaviour, especially, 
as expected, on the asymmetric parts of the diagram. 

%%%%%%%%%%%%%%%%%%%%%%%%%%%%%%%%%%%%%%%%%%%%%%%%
%%%%%%%%%%%%%%%%%%%%%%%%%%%%%%%%%%%%%%%%%%%%%%%%
%%%%%%%%%%%%%%%%%%%%%%%%%%%%%%%%%%%%%%%%%%%%%%%%

\section{Direction of phase separation}\label{section-average}

Let us consider a system whose proton fraction $Z/A$ and total density $\rho$ 
are such that it is situated inside instability region. 
This is illustrated on the right-hand side of fig.\ref{cs-plan} 
for a neutron-rich system ($Z/A=0.3$). 

If equilibrium is reached, this system will undergo 
phase separation according to Gibbs construction. 
We can see that the corresponding phases in coexistence, represented as black dots 
on the coexistence border on 
the right part of  %<<< PC
fig.\ref{cs-plan}, 
do not belong to the line of constant proton fraction.  
%PC $\rho_p/\rho_n=(Z/A)/[1-(Z/A)]$.
%PC , but are situated on each side of this line. 
We can %PC also 
remark that the liquid fraction is closer to symmetric matter than the gas 
fraction: this behavior is a %PC simple 
consequence of the 
symmetry-energy minimization in the dense phase.
%PC l'ordre des deux phrases invers
This inequal repartition of isospin between the two phases is the 
well-known phenomenon of isospin fractionation \cite{Muller-Serot}. 

In order to quantify the phenomenon of isospin fractionation, we will 
% PC >>> 
study %PC use 
the associated 
%This observation can be expressed in terms of 
directions in the density plane. 
The line of constant proton fraction defines a direction $\mathbf{u}_{Z/A}$. Let us introduce the isoscalar direction $\mathbf{u}_{\rho}$, which corresponds to the axis of total density $\rho=\rho_n+\rho_p$. The equilibrium direction $\mathbf{u}_{eq}$ is given by the straight line that joins the couple of phases in coexistence. 
Isospin fractionation is linked to the difference between $\mathbf{u}_{eq}$ and $\mathbf{u}_{Z/A}$. The equilibrium direction being 
%PC shifted 
rotated %<< PC
towards $\mathbf{u}_{\rho}$, phase separation gives a liquid more asymmetric than the gas.

We now turn to study isospin fractionation if phase separation occurs out of equilibrium. 
Specifically, if the system is brought inside the spinodal region in a reaction too fast for 
global %PC
equilibrium to be achieved, its evolution will be driven by local instabilities 
instead of global Gibbs construction. This is spinodal decomposition \cite{Chomaz-PhysRep} : the local properties of the 
constrained %PC
entropy curvature determine the development of a spinodal instability into phase separation. 
The information for this study is contained in the free-energy curvature matrix given by :

\begin{equation}
C=
\left( 
\begin{array}{ll}
\partial^2 f/ \partial \rho _n^2 & \partial^2 f/ \partial \rho _n\partial \rho _p\\ 
\partial^2 f/ \partial \rho _p \partial \rho_n & \partial^2 f/ \partial \rho _p^2\\ 
\end{array}
\right)
=
\left( 
\begin{array}{ll}
\partial\mu _{n}/\partial\rho _{n} & \partial\mu _{p}/\partial\rho _{n}\\ 
\partial\mu _{n}/\partial\rho _{p} & \partial\mu _{p}/\partial\rho _{p}\\ 
%\partial_{\rho _{n}} \mu _{n} & \partial_{\rho _{p}} \mu _{n}\\ 
%\partial_{\rho _{n}} \mu _{p} & \partial_{\rho _{p}} \mu _{p}\\ 
\end{array}
\right)
\end{equation}

\begin{figure}[htbp]
\includegraphics*[width=1\linewidth]{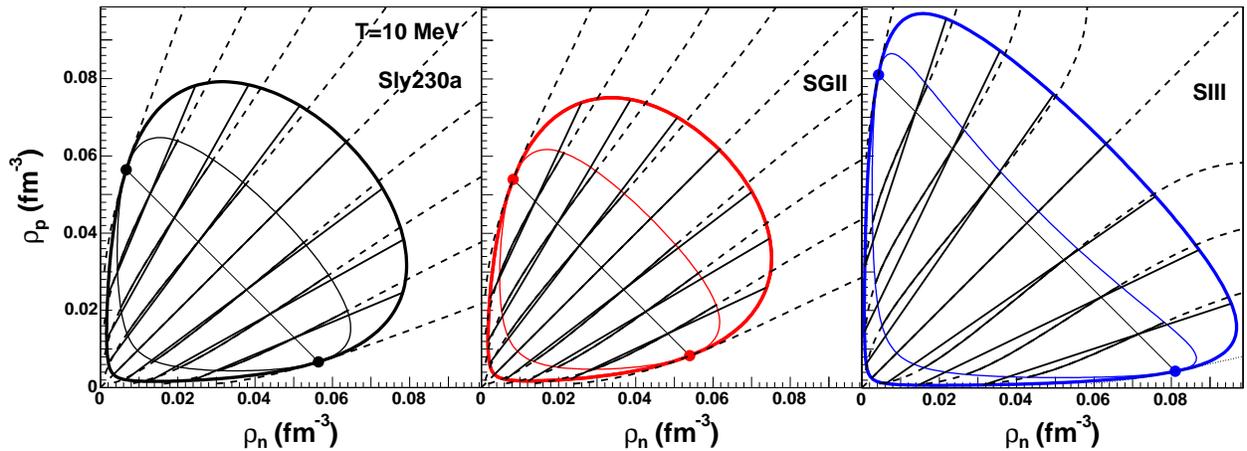}
\caption
	{
	Global representation of phase-separation directions 
	through the density plane, for three different Skyrme parameterizations. 
	Coexistence and spinodal boundaries are represented 
	as in fig.\ref{cs-plan}.
	A few selected equilibrium directions are given by straight 
	lines joining the high and low-density branches of the
	coexistence curve 
	(these straight lines are everywhere tangent 
	to the equilibrium direction $\mathbf{u}_{<}^{eq}$).
	%%%%%
	%The corresponding instability directions $\mathbf{u}_<^{0}$ are represented 
	%for 7 chosen points at a fixed total density and different asymmetries. 
	%%%%%
	Phase-separation directions in spinodal decomposition
	are represented by a selection of lines 
	which are everywhere tangent to the instability direction $\mathbf{u}_{<}^{0}$.
	They are drawn as full lines 
	inside the spinodal region where they correspond to an 
	instability direction, and prolongated in dashed lines ouside 
	the spinodal, where there is no more local instability.
	} 
\label{dir-3F}
\end{figure}

This matrix is defined for each point of the density plane. The lower eigen-value $C_<$ corresponds to the minimal curvature of $f$ at the considered point. The spinodal 
border %PC curve 
is the ensemble of points for which $C_< =0$. 
The instability region is defined by $C_< <0$ : in such points, there are directions of abnormal curvature, corresponding to a concave free energy. 
The eigen-vetctor $\mathbf{u}_<$ associated with the lower eigen-value gives the 
direction of most negative curvature: this is the instability direction. 
It is represented by the 
%PC arrow 
%arrows %PC 
double arrow
on the example of fig.\ref{cs-plan}, 
where we can see that spinodal decomposition leads to a more pronounced fractionation 
than equilibrium,
%PC
the denser phase getting closer to the symmetric matter.
%
%
%cor061009
It is important to remark at this point that the instability direction does not  give a direct quantitative measurement of the final fractionation after phase separation. For this to be true,  the effect of the evolution of density fluctuations on the isospin composition has to be completely neglected. If this is certainly an approximation \cite{ColonnaMatera}, it has been however shown in explicit transport calculations \cite{Chomaz-PhysRep} that the initial instability largely dominates 
the subsequent dynamics. The 
interest of employing this simplified approach as complementary to more sophisticated transport codes is that, within this approximation,  the spinodal mechanism can be directly compared to phase separation at equilibrium. 
Indeed   
%cor061009
%
%
%We can remark here that 
the equilibrium direction can also be expressed as the eigen-vector of a curvature matrix, like the instability direction. 
Let us define $f^{eq}$ as the free energy at equilibrium, 
\emph{i.e.} after correction of the curvature anomaly by phase mixing according to Gibbs construction. This construction imposes a positive curvature at each point, and a zero curvature between two phases in coexistence. As a result, in the coexistence region the lower eigen-value of the $f^{eq}$ curvature matrix is zero, and the associated eigen-vector is what we have 
defined as the equilibrium direction, 
$\mathbf{u}_{eq}$. %PC 
In the following, we will denote $f^{0}$ and $f^{eq}$ the free energy before and after Gibbs construction, their respective curvature eigen-modes being $(C_{<}^{0},\mathbf{u}_{<}^{0},C_{>}^{0},\mathbf{u}_{>}^{0})$ and
$(C_{<}^{eq},\mathbf{u}_{<}^{eq},C_{>}^{eq},\mathbf{u}_{>}^{eq})$. 
A qualitative impression of how equilibrium and instability directions compare through the density plane can be gained from fig.\ref{dir-3F} for the three Skyrme forces. 
%PC >
In order to follow the eigen-direction of the curvature matrix, we draw curves
which are everywhere tangential to the curvature eigenvectors.    
% < PC
Both directions of phase separation
%PC >>>
at equilibrium (Gibbs construction) and out of equilibrium (spinodal decomposition)  
%PC <<<<
remain quite close, and essentially isoscalar \cite{Margueron-isoscalar, Baran-PhysRep410}. A comparable degree 
of isospin fractionation, with formation of a more symmetric liquid, is conserved all through the instability region. The dominant feature is that 
the instability direction leads to more fractionation than equilibrium. 

%
%figure 3
%

In order to have a quantitative comparison of the different fractionations, 
it is convenient to express the directions of phase separation 
as slopes with respect to the isoscalar direction $\delta\rho_3/\delta\rho$.
This quantity gives a measure of the degree of fractionation associated with phase separation.
It can in principle vary from zero 
(in the case of a purely isoscalar phase separation), 
to infinity (in the case of a purely isovector instability,
corresponding to a separation of protons from neutrons into two phases at the same baryon density). In practice $\delta\rho_3/\delta\rho$ is always a small number \cite{Margueron-isoscalar, Baran-PhysRep410}.
This is easy to understand considering that
%PC>>>
i) proton and neutron kinetic energies favour equal-size Fermi spheres
and ii)
%PC <<<< 
the nuclear interaction is more attractive in the proton-neutron 
than the proton-proton and neutron-neutron channels 
in any realistic nuclear physics model ($T=0$ contribution), which 
favours the existence of a mixed isospin bound phase. 
%PC 
%PC The isoscalar character of the transition can also be understood 
%PC in thermodynamic words. Indeed the nuclear liquid-gas phase 
%PC transition is known to have
%PC a scalar order parameter like ordinary liquid-gaz, namely
%PC the total baryon density. 
%PC A scalar order parameter implies that there is a single instability direction 
%PC (i.e. a unique negative eigenvalue at most) in the nuclear entropy surface. 
%PC The fact that this order parameter can be associated to the density means 
%PC that this direction must have a dominant projection along the isoscalar 
%PC (i.e. total baryon density) axis. 
%PC This rules out the possibility of isovector instabilities.
 
%Let us define the isovector density $\rho_3=\rho_n-\rho_p$. This quantity is read on the %isovector axis, which is perpendicular to the isoscalar axis giving the total density. We %have chosen to represent a direction by the corresponding slope $\delta\rho_3/\delta\rho$, %which is zero for isoscalar direction.

%figure 4
%
\begin{figure}[htbp]
%PC \includegraphics*[height=0.65\linewidth]{c_pan_dr3sdr_T10_v1_tl.eps}
\includegraphics*[width=1\linewidth]{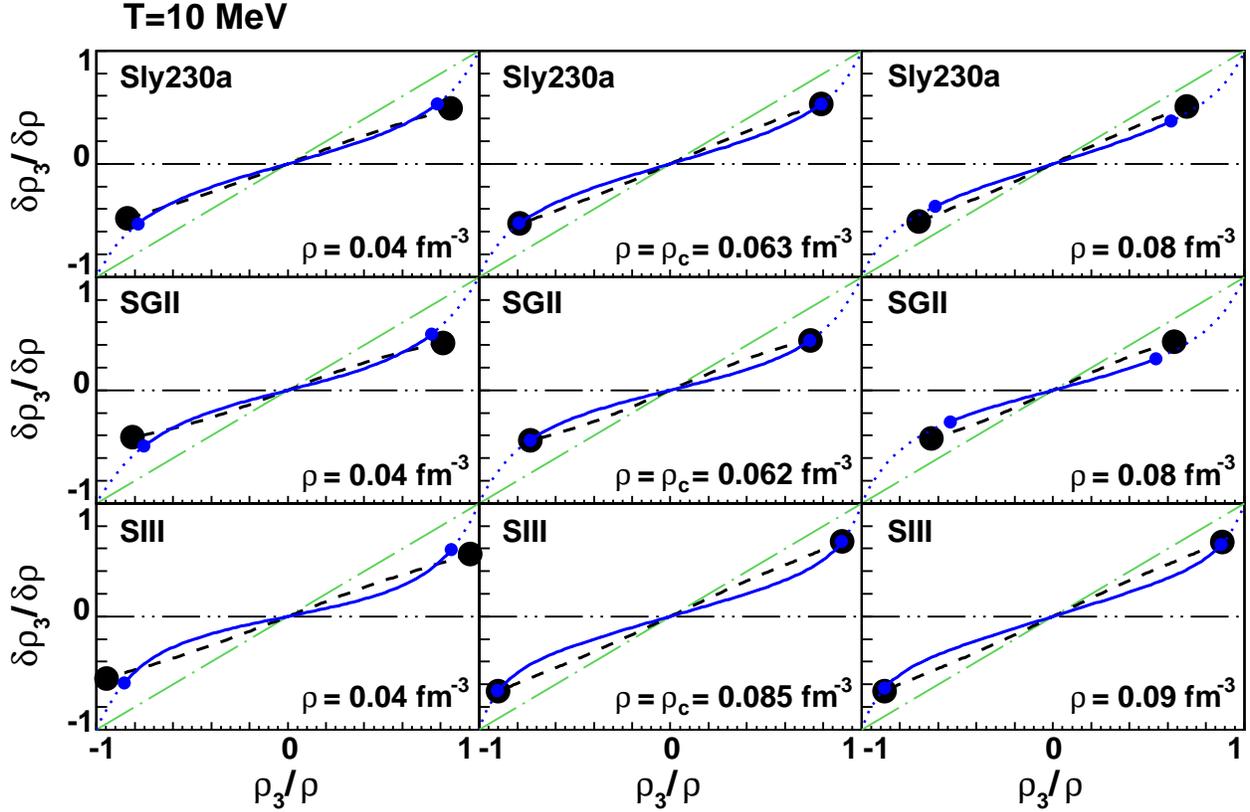}
\caption
	{
	%PC >>>>
	Phase-separation directions 
	$\delta \rho_3 / \delta \rho$
	at equilibrium (dash)  and in spinodal decomposition (solid) 
	%PC , expressed as the 
	%PC  slope with respect to the isoscalar direction (see text). 
	%PC They are represented 
	as a function of the asymmetry $\rho_3/\rho$ 
	for three different Skyrme parameterizations at different values of $\rho$ :
	 %$\rho=cst$, with $cst <,=,>\rho_c$, where $\rho_c$ is the critical density.
	below, at and above the associated critical density. The instability direction 
	is continued by the lowest curvature direction (dotted line).
	The dot-dash straight line gives the direction of constant $Z/A$ while 
	the double-dot-dash horizontal line gives the isoscalar direction 
	$\delta \rho_3 = 0$. 
	%PC The instability direction is represented as a full line, 
	%it belongs to the line of $\mathbf{u}_<^{0}$ direction 
	%which is prolongated in dotted line.
	 %PC the dashed line %existing between the critical points 
	 %PC gives the equilibrium direction.
	Small dots give spinodal boundaries %(for instability direction) 
	and large ones coexistence region boundaries.
	%PC, respectively. %(for equilibrium direction).
	%PC <<<<
	} 
\label{dir-pan}
\end{figure}

The evolution of $\delta \rho_3/\delta \rho$ (degree of fractionation)
with $\rho_3/\rho$ (system asymmetry)
is given by fig.\ref{dir-pan} for three different total densities :
below, at, and above the respective critical density $\rho_c$
(\emph{i.e} the total density at critical points). 
In this representation, the direction of constant 
proton fraction is a diagonal straight line while the isoscalar 
direction is an horizontal line.
On this figure, equilibrium and instability direction are compared 
for the different Skyrme parameterizations.

Phase-separation evolutions verify some expected constraints. 
First, the isospin invariance of Skyrme interactions
%invariance respect to neutron-proton exchange 
ensures that the graphs are symmetric %have an odd parity 
with respect to $\rho_3=0$. 
It also imposes that all directions become purely 
isoscalar in the case of symmetric matter, i.e. all curves must join 
at 
the origin $(\delta\rho_3/\delta\rho, \rho_3/\rho)=(0,0)$. %PC <<<<
Another constraint is given by the ending points of the curves.
%In figure \ref{dir-pan} the direction of $\mathbf{u}_<$ is represented 
%as a full line , although it can be interpreted as the 
%instability direction only inside the spinodal region, it can be defined 
%for any point of the density plane. We have prolongated its representation 
%with a dotted line until the physical borders of this plane $\rho_3=\pm \rho$, 
%corresponding to pure neutron and pure proton matter. 
At these borders ($|\rho_3|=\rho$), the vanishing of one species means that its chemical potential at finite temperature goes to $- \infty$, implying an infinite curvature of the free energy in the corresponding direction.
The minimal curvature $\mathbf{u}_<^0$ is then oriented along the axis of the remaining species, which is also the direction of constant $Z/A$. 
%Indeed,   is then forced to the perpendicular direction, corresponding to the 
%axis of the remaining species. It is an additional constraint on the shape of 
%the graph obtained for instability direction that it belongs to a curve that 
%ends joining the line of constant $Z/A$ direction.

%PC >>>>
The first observation is that spinodal and phase-equilibrium directions
are very close and lie between the isoscalar direction and the constant-proton-fraction one. 
This means that they lead to a rather similar fractionation. 
In order to get a deeper insight, let us consider the various densities in more details.  
%PC <<<<

Let us consider the case of the axis at $\rho=\rho_c$, for which the instability region ends at the critical points. Between these extremities, the instability direction induces more fractionation than equilibrium
since %its slope is smaller. %PC <<<
it is closer to isoscalar direction.
At the critical points, where the spinodal and coexistence curves are tangent, both directions of phase separation join 
to become tangential to the coexistence and spinodal borders. %PC <<< 
This is an additional constraint on phase-separation directions  %PC <<< 
which, with the already discussed constraints, explains the  %PC <<< 
similarities between the phase-separation directions at and out of %PC <<< 
equilibrium.  %PC <<<     
%There is no such boundary condition for other axis at constant 
%density $\rho \neq \rho_c$. 
%This is not the case for the other densities. 
This additional constraint does not apply 
if the total density we consider is not $\rho_c$. 
We observe that for $\rho<\rho_c$, a small inversion occurs around the extremities of the instability region. %, as noticed above. 
For $\rho>\rho_c$ instead, 
%the directions do not join within instability region and the instability 
%direction is the closest to isoscalar direction.
a phase separation out of equilibrium induces more fractionation 
for all isospin asymmetries.

These observations stand for the three Skyrme forces that we have considered. The quantitative differences, which essentially distinguish SIII from the two more recent parameterizations, do not affect the qualitative features that have been described. 
%PC >>>>
This is partly due to the above discussed constraints. 
%PC <<<<

%%%%%%%%%%%%%%%%%%%%%%%%%%%%%%%%%%%%%%%%%%%%%%%%
%%%%%%%%%%%%%%%%%%%%%%%%%%%%%%%%%%%%%%%%%%%%%%%%
%%%%%%%%%%%%%%%%%%%%%%%%%%%%%%%%%%%%%%%%%%%%%%%%
\section{Fractionation and isocaling observables}\label{section-isoscaling}

%Isoscaling is an experimental observation 
%on the properties of fragment isotopic yields.
In a variety of experimental situations,
it has been observed that the
isotopic-yield ratio for two reactions 
(denoted $(1)$ and $(2)$) %PC <<<<
of different asymmetry
approximatly follows an exponential form \cite{Tsang}:
\begin{equation}
R_{iso}^{(1),(2)}(N,Z)=Y^{(2)}(N,Z)/Y^{(1)}(N,Z) \propto exp(\alpha_n N+\alpha_p Z).
\end{equation}
This phenomenon is known under the name of isoscaling.
In an equilibrium interpretation in the grand-canonical ensemble,
the isoscaling parameters $\alpha_q$ should be linked to the chemical
potentials and temperature of the systems, 
$\alpha_q=(\mu_q^{(2)}-\mu_q^{(1)})/T$.
It has been proposed to extract from the measured isoscaling parameters 
the symmetry-energy coefficient $C_{sym}$ 
through the approximate formula \cite{Ono,Shetty}: 
\begin{equation}
\alpha_n=4C_{sym}
%PC /T 
([Z/\langle A\rangle ]_{(1)}^2-[Z/\langle A \rangle]_{(2)}^2)
/T %PC
.
\label{alpha_1}
\end{equation}
This relation, derived for the grand-canonical ensemble in the 
saddle-point approximation \cite{Ono}, holds in the hypothesis of equilibrium
at the time of fragment formation. This means that to extract $C_{sym}(Z)$ from 
the measured isoscaling parameters, modifications of $\alpha_n$ due to secondary
decay have to be accounted for and the proton fraction of the most probable 
isobar for each $Z$ should be known at the time of fragment formation .

Because of the small variation of $\alpha_n$ with the atomic number,
an average estimation of $C_{sym}$ can be obtained using for
$[ Z/\langle A\rangle ]_{(i)}$ the average proton content of the 
liquid fraction \cite{Ono}.
Since even this quantity is not experimentally known, 
a similar formula can be found in the literature, 
where, however, the proton fraction 
of the fragmenting source 
$[Z_0/A_0 ]_{(i)}$ %PC <<<<
is used 
instead of $[Z/\langle A\rangle ]_{(i)}$:
%leading to %PC <<<<
this leads to a quantity we call
$C_{sym}^0$ %PC <<<<
\cite{Tsang2,LeFevre,Botvina}.
%
%PC \begin{equation}
%PC \alpha_n=4C_{sym}^0/T ([Z_0/A_0 ]_{(1)}^2-[Z_0/ A_0]_{(2)}^2).
%PC \label{alpha_2}
%PC \end{equation}
% 
This approximation is supported by the fact that 
in the zero-temperature limit the proton fraction of the 
fragments converges to the proton fraction of the source \cite{Botvina}.

%In the general case however 
However, in general
this does not need to 
be the case and the approximated $C_{sym}^0$ %PC from eq.(\ref{alpha_2})
may differ from the value $C_{sym}$. %PC obtained with eq.(\ref{alpha_1}). 
In particular, both in the equilibrium hypothesis 
at finite temperature and in the case of dynamical fragment formation through
spinodal decomposition, as we have seen, we can expect the phenomenon
of fractionation: fragments should not have the same
isospin ratio as the fragmenting source. This should
affect the value of $C_{sym}$ obtained from the isoscaling coefficients. 

Our equilibrium nuclear-matter approach does not allow to give quantitative 
predictions on the actual isospin content of fragments obtained 
out of the different phase-separation mechanisms; we can however make some qualitative
order of magnitude estimations of the possible deformation that isospin 
fractionation may induce on symmetry-energy measurements.
 
Dynamical transport models \cite{Baran-PhysRep410,Ono} indicate that 
clusterization takes place at a relatively well-defined density
inside the spinodal region determined by the collision dynamics and
corresponding to the first development of instabilities, 
for which we take a typical value around $\rho_0/3$.
Spinodal decomposition is then a phenomenon fast enough 
for fragment formation dynamics to be dominated 
by the amplification of the most unstable modes 
at this density \cite{Chomaz-PhysRep}. 
This density then determines the phase-separation direction.

%
% fig.5
%
\begin{figure}
\includegraphics*[width=1\linewidth]{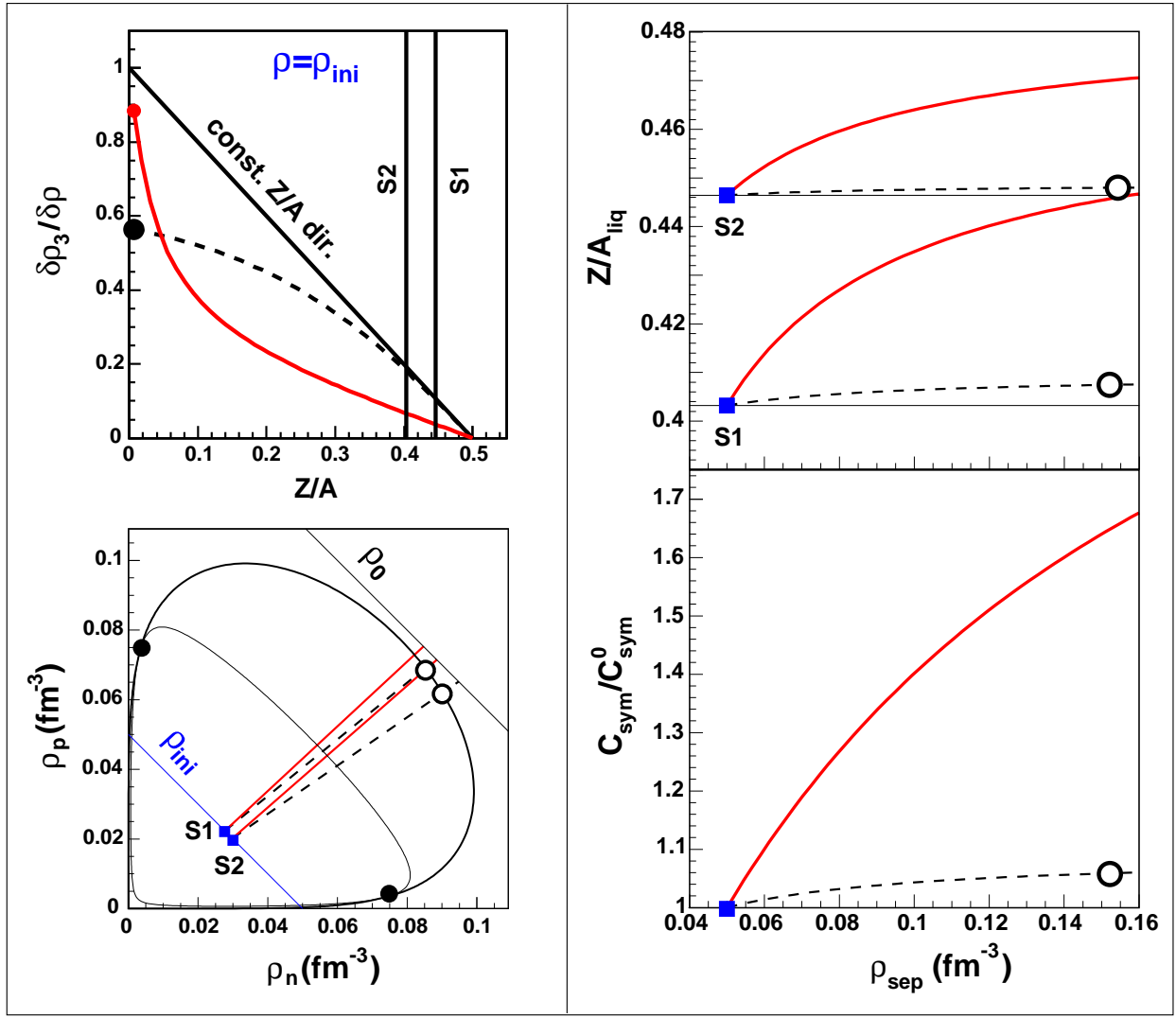}
\caption
	{
	Influence of isospin fractionation on $C_{sym}$ measurement at $T=4 MeV$. 
	Left side: 
	phase-separation directions.
	Top-left: 
	separation direction at a typical density of fragment 
	formation $\rho=\rho_{ini}=0.05 fm^{-3}$,
	as a function of the global proton fraction 
	for equilibrium (dashed line) and spinodal instability (full line). 
	Vertical lines indicate the two systems S1$\sim$$^{112}$Sn, S2$\sim$$^{124}$Sn.
	Bottom-left:
	representation of the corresponding phase separations in the density plane.
	Right side:
	consequence on symmetry-energy measurement.
	Top-right :
	average isotopic composition of the liquid-like fraction 
	as a function of fragment density at separation,
	for the two systems S1 and S2.
	Bottom-right:
	Symmetry energy coefficient $C_{sym}$ compared to the approximate expression 
	obtained neglecting fractionation $C_{sym}^0$, as a function of separation density,
	for equilibrium (dashed line) and spinodal instability (full line).
	In the case of equilibrium, the liquid position is fixed in the density plane:
	it is indicated by the empty circles. 
	%indicate the resulting proton fractions and $C_{sym}/C_{sym}^0$ ratio.
	}
\label{fig05} % optional figure label, must be unique
\end{figure}

The left part of fig.\ref{fig05} displays, similar to fig.\ref{dir-pan}
above, the directions of phase equilibrium and
spinodal decomposition as a function of Z/A at a typical temperature 
of $T=4 MeV$. The case of two systems with global 
proton fractions corresponding to $^{112}$Sn and $^{124}$Sn is shown
by the vertical lines. 
We can see that fractionation at equilibrium is negligible in agreement 
with ref.\cite{Botvina}, and fragmentation
of more exotic nuclei should be studied to explore this effect.
However even for these stable systems, spinodal decomposition  
gives an important fractionation,
and the effect would be amplified using more asymmetric systems.

Once the direction of phase separation is known, 
the final composition of the fragments depends only on their density 
at freeze-out, when they become separated. 
We call this density $\rho_{sep}$. 
The average $Z/A$ of the liquid fraction is represented in the central
part of fig.\ref{fig05} as a function of the density at separation.
This density must be between the initial density and the saturation value, but its 
actual value is not known from fundamental principles. 
It is generally assumed that fragments at separation are 
at normal density $\rho_{sep}\approx \rho_0$. Recent AMD calculations
however seem to indicate that lower values 
$\rho\approx 0.1 fm^{-3}$ may be more realistic \cite{Shetty,Ono2}.

%It determines the liquid proton fraction. 
The  asymmetry coefficient $C_{sym}$ 
deduced through eq.(\ref{alpha_1})
is compared to its approximate value $C_{sym}^{0}$ 
obtained with the proton fraction of the fragmenting source
%eq.(\ref{alpha_1}) 
in the right part of fig.\ref{fig05}. 
%considering a range from $\rho_{ini}$ to $\rho_0$. 
Little change 
(few \% at maximum) %PC <<<< 
is observed in the case of equilibrium,
again consistently with the predictions 
of the statistical model \cite{Botvina}. 
However, the stronger fractionation effect in the 
case of spinodal decomposition 
sensitively %PC increases
decreases, %PC <<<<
by up to $50 \%$ (or more), %PC <<<<
the 
approximate %PC <<<<
asymmetry coefficient.
This implies that the low values reported for $C_{sym}$
in ref.\cite{LeFevre} may be alternatively interpreted 
as a signal of isospin fractionation
in a spinodal decomposition. %PC <<<< 
%PC , if fragmentation occurs out of equilibrium.

%%%%%%%%%%%%%%%%%%%%%%%%%%%%%%%%%%%%%%%%%%%%%%%%
%%%%%%%%%%%%%%%%%%%%%%%%%%%%%%%%%%%%%%%%%%%%%%%%
%%%%%%%%%%%%%%%%%%%%%%%%%%%%%%%%%%%%%%%%%%%%%%%%
\section{Isospin fluctuations}

In section \ref{section-average}, we have studied the direction of phase separation, corresponding to the direction of minimal curvature of the free energy $\mathbf{u}_<$.
This direction can be qualitatively linked to the average isospin content of fragments
issuing from the phase-separation mechanism. As we have seen in section \ref{section-isoscaling}, some quantitative predictions can also be obtained if some 
extra hypotheses are %PC assumed
made, %PC <<<
namely that fragment-formation dynamics is dominated by an
amplification of instabilities from well-defined initial density and temperature. 
Within the same line of reasoning,
 the other eigen-vector $\mathbf{u}_>$ giving the direction of maximal curvature 
 can be linked to fluctuations in the system composition, that we now turn to analyze.
 
A physical system characterized by a given value of the isoscalar density $\rho$ 
at equilibrium presents a distribution of $\rho_3$ given by:

\begin{equation}
P_{\beta,\mu,\mu_3}(\rho_3) \propto e^{V(s_c^{\beta,\mu}(\rho_3)+\beta \frac{\mu_3}{2} \rho_3)}
\label{distribution}
\end{equation}
where $\mu_3=\mu_n-\mu_p$ is the isovector chemical potential,
and $\mu=\mu_n+\mu_p$ the isoscalar chemical potential.
The constrained entropy $s_c^{\beta,\mu}(\rho_3)$
is defined by the %PC Lagrange 
Legendre %PC <<<<
transform (\emph{cf} eq.(\ref{s_c})):
\begin{equation}
s_c^{\beta,\mu}(\rho_3)=s(e,\rho,\rho_3)-\beta e+\beta \frac{\mu}{2} \rho
\end{equation}
%
%Fluctuations are linked to the free-energy curvature, as long as this curvature is strictly positive. Let us consider a system for which we study the observable $\hat A$, controlled by the Lagrange parameter $\lambda$. The probability distribution of A is then related to the entropy as :

%\begin{equation}
%P(A) \propto exp(S(A)-\lambda A)
%\end{equation}

In the case of an infinite system, the distribution (\ref{distribution}) is 
characterized by a unique value, $P(\rho_3)\propto \delta(\rho_3-\rho_3^0)$,
given by the maximum $\rho_3^0$ of the function $s_c^{\beta,\mu}(\rho_3)+\beta \frac{\mu_3}{2} \rho_3$.
Indeed the correspondence between densities and chemical potentials is one-to-one
for infinite systems
since fluctuations go to zero at the thermodynamical limit. %PC <<<<
If conversely we consider systems of finite volume V, 
%PC (assuming it is big enough for thermodynamic approximation to apply), 
this distribution can be approximated 
by a gaussian
if the system is big enough: 
%PC <<<< 
%The curvature properties of $lnP(A)$ are those of $S(A)$.
%In a gaussian approximation, this distribution can be expressed as :
%\begin{equation}
%P(A) \propto exp \left[ (A-A_0)^2 / (2\sigma_A^2) \right]
%\end{equation}
%
\begin{equation}
P_{\beta,\mu,\mu_3}(\rho_3) \propto e^{-\frac{(\rho_3-\rho_3^0)^2}{2\sigma_{\rho_3}^2}}
\label{gaussian}
\end{equation}
where the distribution variance 
%$\sigma_A^2=<\hat A^2>-<\hat A>^2$ is representative of the fluctuations of $A$ around its mean value $A_0=<\hat A>$.
$\sigma_{\rho_3}^2=<\rho_3^2>-<\rho_3>^2$, which measures the isospin fluctuation,
is directly linked to the entropy curvature. Indeed if the entropy has a finite
curvature in $\rho_3$ direction, 
a development of 
%$S(A)-\lambda A$ 
$s_c^{\beta,\mu}(\rho_3)$
around %$A_0$ 
$\rho_3^0$ gives:
\begin{equation}
s_c^{\beta,\mu}\left(\rho_3\right)+\beta\frac{\mu_3}{2}\rho_3 = \left[s_c^{\beta,\mu}(\rho_3^0)+\beta\frac{\mu_3}{2}\rho_3^0\right] + \frac{\left(\rho_3-\rho_3^0\right)^2}{2} \frac{\partial^2 s_c^{\beta,\mu}}{\partial \rho_3^2} |_{\rho_3^0}
\label{expansion}
\end{equation}
%
%permits to link $\sigma_A^2$ to the entropy curvature :
%\begin{equation}
%S(A)-\lambda A = [S(A_0)-\lambda A_0] + \frac{(A-A_0)^2}{2} \left( \frac{\partial^2 S}{\partial A^2} \right)_{A_0}
%\end{equation}
%
%with:
%\begin{equation}
%\sigma_A^2= \left( \frac{\partial^2 S}{\partial A^2} \right)^{-1}_{A_0}
%\end{equation}
so that
\begin{equation}
\sigma_{\rho_3}^2= -\frac{1}{V}
\left( \frac{\partial^2 s_c^{\beta,\mu}}{\partial \rho_3^2} \right)_{\rho_3^0}^{-1}=
\frac{1}{\beta V}
\left( \frac{\partial^2 f}{\partial \rho_3^2} \right)^{-1}_{\rho_3^0}
%
%cor061009
%=\frac{\rho T}{2VC_{sym}(\rho)}
%cor061009
%
\label{fluct}
\end{equation}
This is related to the symmetry-energy coefficient $C_{sym}$ defined from the free-energy curvature by $\partial^2 f/\partial \rho_3^2=\frac{2}{\rho}C_{sym}(\rho)$, which is independent of $\rho_3$ in the parabolic approximation \cite{baoanli} :
\begin{equation}
\sigma_{\rho_3}^2=\frac{\rho}{2\beta V C_{sym}(\rho)}
\label{fluct-Csym}
\end{equation}
%
%
%cor061009
%where the symmetry energy
%$C_{sym}(\rho)=\frac{\rho}{2}\partial^2 f/\partial \rho_3^2$
%is independent of $\rho_3$ in the parabolic approximation \cite{baoanli}.
%cor061009
%
The neglect of higher order terms in eq.(\ref{expansion}) is not fully justified for a genuinely finite system, where we expect deviations from the gaussian ansatz
(\ref{gaussian});
it is however consistent with our 
%PC thermodynamical 
mean-field %PC <<<
approximation.
%
%cor061009
Equations (\ref{gaussian}), (\ref{fluct-Csym}) show that isospin distributions are directly linked to the symmetry properties of the effective interaction. This is why isotopic 
distributions measured in fragmentation reactions have been studied by many authors and their widths tentatively connected to the symmetry-energy coefficient $C_{sym}$ at subsaturation density 
\cite{Botvina,Liu}.
%cor061009
%

%Such relation stands only if $P(A)$ can be approximated by a gaussian distribution : a necessary condition is to have a strictly positive curvature of the entropy.

%At thermodynamic limit, the system is described per unit volume : $S(A)=Vs(a)$ with $a=A/V$. The relation between $s$ curvature and $a$ variance the becomes :
%\begin{equation}
%\sigma_a^2= \left( \frac{\partial^2 s}{\partial a^2} \right)^{-1}_{a_0}/V
%\end{equation}

%In the case of an infinite system, there is no fluctuation although $s$ curvature is finite, because of the factor $1/V$. Considering system of volume $V$ (assuming it is big enough for thermodynamic approximation to apply), the fluctuations are proportional to $\left( \frac{\partial^2 s}{\partial a^2} \right)^{-1}_{a_0}$.
 
Let us now consider a system at equilibrium inside the spinodal region at a point 
$(\rho_n,\rho_p)$. Since this situation is unstable for an homogeneous system, it will
separate into a liquid and a gas fraction. As we have discussed in section \ref{section-average}, 
this separation follows the direction of minimal curvature $\mathbf{u}_<$ of the 
free energy. 
%This direction, together with the orhogonal eigen-vector $\mathbf{u}_>$   
%defines 
%For each point of the density plane, the eigen-directions $\mathbf{u}_<$, $\mathbf{u}_>$
The directions of the two orthogonal eigenvectors $\mathbf{u}_<$ and $\mathbf{u}_>$
define a set of orthogonal eigen-observables $(\rho_<,\rho_>)$ corresponding to a rotation of the set $(\rho_n,\rho_p)$ with an angle $\alpha=(\mathbf{u}_<,\mathbf{u}_n)$:  
%In the following, we will note these eigen-observables , where 

\begin{equation}
\begin{array}{l}
\rho_< =\rho_n cos\alpha+\rho_p sin\alpha \\
\rho_> =\rho_n sin\alpha-\rho_p cos\alpha
\end{array}
\end{equation}

At each point of the system evolution along the axis $\rho_<$, 
the inverse curvature along the orthogonal axis measures the fluctuation of 
$\rho_>$ . 
Since the direction of phase separation $\mathbf{u}_<$ is essentially isoscalar 
($\alpha\approx \pi/4$), 
the observable $\rho_<$ is close to the total system density
while %the direction of 
$\rho_> \approx \frac{\rho_n-\rho_p}{\sqrt{2}}=\frac{\rho_3}{\sqrt{2}}$
approximately gives the isospin content of the system. 
In that sense, the fluctuations of $\rho_>$ are related to the isospin fluctuations of the system. The actual isospin fluctuations can be calculated projecting the fluctuations of $\rho_>$ on the $\rho_3$ direction.

If the process occurs at equilibrium, phase separation follows
Gibbs construction rules, which constrain the two phases on the coexistence curve.
Thus, the physical equilibrium fluctuations for the system 
are associated with the curvature of the free-energy surface 
$f^{eq}(\rho_n,\rho_p)$ in the direction $\rho_{coL}$ ($\rho_{coG}$), given by the 
tangent to the coexistence curve at the liquid (gas) point. 

Since phase separation is a simple straight line, these final fluctuations are nothing 
but the extrapolation to the coexistence border of the $f^{eq}$ inverse curvature  in the $\mathbf{u}_>^{eq}$ direction, calculated at the initial point $(\rho_n,\rho_p)$ .

\begin{figure}[htbp]
\includegraphics*[width=1\linewidth]{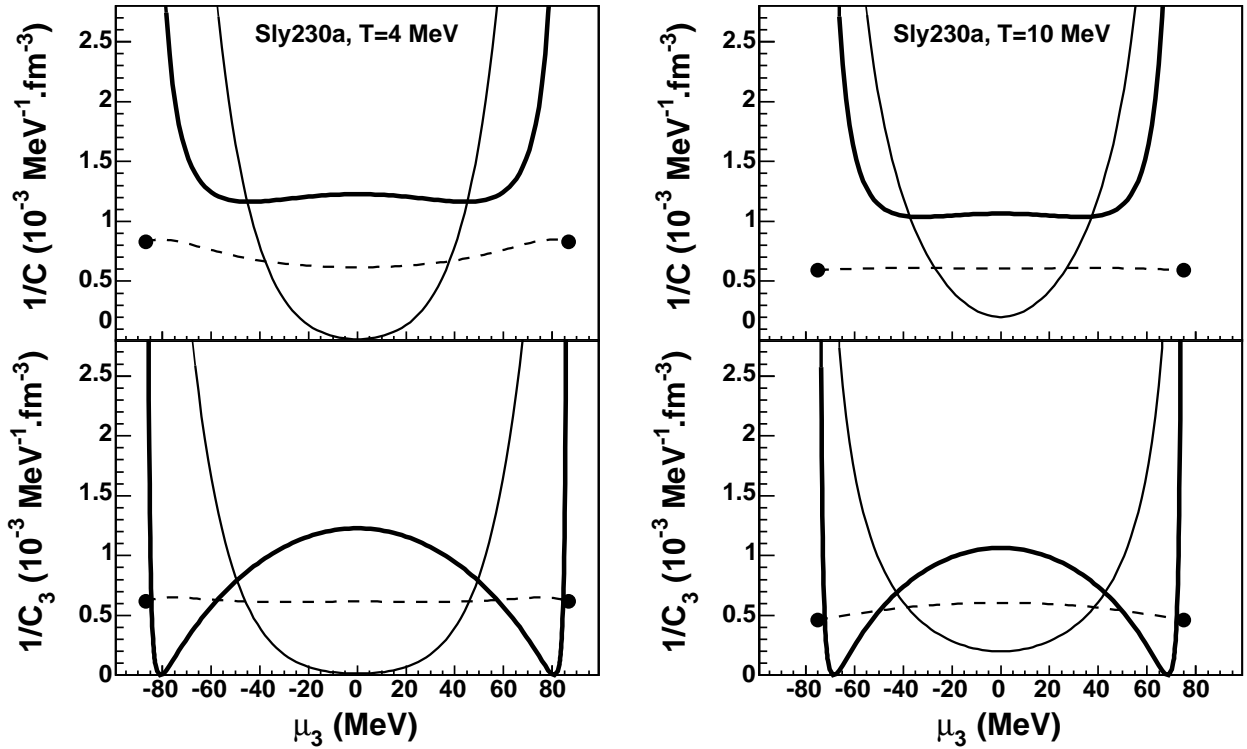}
\caption
	{
	Inverse curvature of the equilibrated free energy $f^{eq}$
	as a function of the isovector chemical potential $\mu_3$. 
	Dashed lines: curvature taken in the direction $\mathbf{u}_>^{eq}$ for points inside 
	the coexistence region along the axis $\rho=\rho_c$.
	Thick (thin) lines: curvature taken in the direction $\mathbf{u}_{coL}$ ($\mathbf{u}_{coG}$)
	tangent to the liquid (gas) branch of the coexistence 
	curve. Left part: $T=4 MeV$. Right part: $T=10 MeV$.
	Lower part: projection on $\rho_3$ axis of the different curvatures considered.
	}
\label{fluc-pan}
\end{figure}

%The free-energy surface $f^{eq}(\rho_n,\rho_p)$ is obtained by Gibbs 
% construction. At each point $(\rho_n,\rho_p)$ the highest 
% eigenvalue of the curvature matrix, i.e. the curvature 
% of $f^{eq}$ in the $\rho_>$ direction, is a measure through eq.(\ref{fluct})
% of the isospin fluctuation 
% , $f^{eq}$ curvature matrix has eigen-vectors 
% $(\mathbf{u}_<^{eq},\mathbf{u}_>^{eq})$ defining 
% the eigen-set of observables $(\rho_<^{eq}, \rho_>^{eq})$. We fix the 
% variable $\rho_<{eq}$ and consider $\rho_>^{eq}$ fluctuations, 
% which are related to $f^{eq}$ curvature in $\mathbf{u}_>$ direction 
% (the free-energy having the same curvature than 
% the entropy). Such fluctuations in the global-system 
% composition correspond to fluctuations in the liquid and gas at 
% equilibrium, with the constraint that each phase is 
% confined to the coexistence line. Thus, while the 
% global system presents $\rho_>^{eq}$ fluctuations, 
% the liquid (gas) has fluctuations of the observable $\rho_{coL}$ 
% ($\rho_{coG}$), corresponding to the direction tangent 
% to the coexistence curve at the liquid (gas) point. 
% In general, the three observables $\rho_>$, $\rho_{coL}$ 
% and $\rho_{coG}$ are different, except for 
% asymmetric matter where all are equal to $\rho_3/
% \sqrt 2$, \emph{i.e.} proportional to the isovector density. 
% The resulting isospin fluctuations are the projection 
% on $\rho_3$ direction of the different fluctuations obtained.

The different curvatures associated with the two phases in coexistence are represented in fig.\ref{fluc-pan} as a function of the isovector chemical potential $\mu_3=\mu_n-\mu_p$, in order to have a common abscissa for liquid and gas.
The projections in the $\rho_3$ direction are given in the lower part of the figure.
The curvature of $f^{eq}$ is almost constant inside the coexistence zone (dashed lines).
Because of the characteristic shape of the coexistence curve (\emph{cf} fig.\ref{dir-3F}) this curvature
is reduced (increased) at the liquid (gas) border. 
It is represented for both phases in full lines.
This very general result is only due to the free-energy structure of the two phases, and we may expect it to stay qualitatively valid independently of finite-size effects and 
phase-separation mechanism.
The extreme values of $\mu_3$ correspond to the critical points, which are defined by a zero curvature along the coexistence curve. 
At these points, 
both curvatures in $\rho_{coL}$ and $\rho_{coG}$ become zero, 
fourth order terms have to be considered in the expansion (\ref{expansion}), 
and the gaussian approximation (\ref{gaussian}) 
breaks down. Since the coexistence curve is closed,  there is a value of asymmetry 
for which the coexistence line is tangent 
to the isoscalar direction, leading to a divergence of the $\rho_3$-projected curvature
on the liquid side. 
However, this still happens  close to the critical point, %PC <<<<
where the gaussian approximation is not enough. %PC <<<< 

Far from the critical point, 
%PC where the gaussian approximation is justified, 
these  inverse  curvatures can be associated with a fluctuation 
through eq.(\ref{fluct}). These fluctuations are largely independent of 
the system temperature inside the coexistence zone. 
They can be 
strictly %PC <<<<
interpreted as isospin fluctuations only for symmetric matter ($\mu_3=0$),
where the $\rho_{coL}$ and $\rho_{coG}$ directions are purely isovector.
For symmetric matter, the liquid fluctuation appears much  %PC <<<<
larger than the gas one.  %PC <<<<
Because of the actual shape of the coexistence line, %PC <<<<
%PC they contain also an isoscalar component 
%PC as shown in the lower part of fig.\ref{fluc-pan}: 
the isospin part of the 
liquid %PC <<<<
fluctuation is a rapidly decreasing function of the global asymmetry,
as shown in the lower part of fig.\ref{fluc-pan}. %PC <<<<
Conversely,  the gas branch presents a fluctuation increasing with the 
asymmetry.   

As far as physical fragmenting systems are concerned,
fluctuations are affected by conservation laws 
which are not accounted for in this grand-canonical approach :
therefore,
%PC <<<
the fluctuations represented in the lower part 
of fig.\ref{fluc-pan} 
%PC cannot be quantitatively 
can only be qualitatively %PC <<<<
interpreted as widths of fragment
isotopic distributions.
%because conservation law constraints
%are %PC <<<<
%not considered in this grandcanonical treatment.

%PC , play an essential role in finite systems fragmentation. 
%PC However
%PC we may expect that the qualitative trend observed in fig.\ref{fluc-pan}
%PC should be independent of the introduction of finite size effects. 
Let us now focus on the difference between %PC <<<
equilibrium fluctuations and those expected in spinodal %PC <<<
decomposition. In both cases we will use the same physical picture %PC <<< 
so that results can be qualitatively compared. %PC <<<<

In the preceeding discussion we have assumed that phase separation is entirely 
determined by equilibrium rules. If, out of equilibrium, spinodal decomposition
is the dominant mechanism, fluctuations in the two dimensional 
$(\rho_n,\rho_p)$ plane grow in time from the instability region until they cause 
the decomposition of the system. Then isospin fluctuations should be calculated as 
a dynamical variable given by the projection over the isovector axis of the 
two dimensional fluctuations varying in time. This approach is followed in numerical codes
\cite{Baran-PhysRep410,Liu},
as well as in simplified analytical  models like in ref.\cite{ColonnaMatera}, which follows the time 
evolution of unstable modes under the influence of a stochastic mean field in the linear
approximation. Both approaches give similar predictions for the isotopic
widths \cite{ColonnaMatera}, 
and suggest narrower distributions than in experimental data and in 
the case of fragmentation in statistical equilibrium \cite{Liu}. 

In these approaches, the initial condition is given by statistical equilibrium 
inside spinodal region of the observables which are not 
associated with the instability.
%, but the associated initial fluctuation is neglected. 
The initial fluctuation of these observables is neglected.
However, 
%PC in the case of spinodal decomposition starting 
%PC from an equilibrated situation, to achieve phase separation 
%PC the system follows the instability direction $\mathbf{u}_<^0$, and  
the inverse curvature along the %PC orthogonal 
axis 
orthogonal to the instability direction $\mathbf{u}_<^0$ %PC <<<< 
is never zero, giving rise to
an initial fluctuation of $\rho_>$ .
%we study the free-energy surface $f^{0}(\rho_n,\rho_p)$, which has at each point the eigen-vectors $(\mathbf{u}_<^{0},\mathbf{u}_>^{0})$ defining the eigen-set of observables $(\rho_<^{0}, \rho_>^{0})$. Fixing the observable $\rho_<^{0}$ that corresponds to the direction of phase separation, we observe fluctuations of $\rho_>^{0}$. 
%
We now want to estimate this effect.
Consistently with
%Similar to 
the reasoning of section \ref{section-isoscaling} above, we assume that this fluctuation is mainly determined at the onset of phase separation, 
which is justified in ref. %PC <<<
\cite{Wen}.
%affecting each future phase in the same way, in contrast with the case of phase equilibrium.

\begin{figure}[htbp]
\includegraphics*[width=1\linewidth]{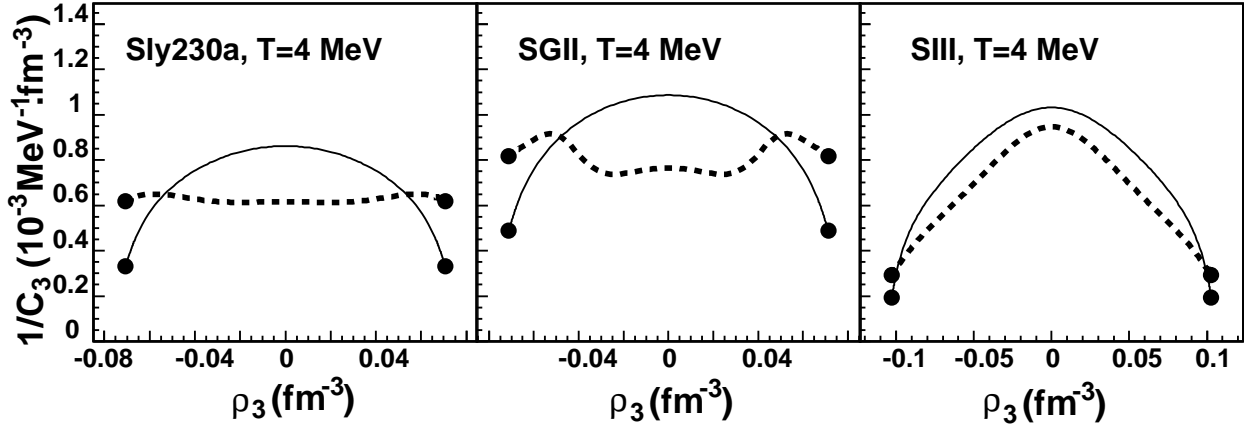}
%nous avons aussi �disposition T=10 MeV
%\includegraphics*[height=0.28\linewidth]{c_invC3_r3rc_3F_t10.eps}
\caption
	{
	Study of isospin fluctuations, compared at equilibrium and for spinodal decomposition 
	for three different Skyrme parameterizations as a function of the total isovector density. 
	The represented quantity is the inverse curvature of the free energy 
	taken in the relevant direction and projected on $\rho_3$ axis.
	Global systems are taken along the axis $\rho=\rho_c$.  
	The system undergoing spinodal decomposition is represented by thin lines 
	(curvature of $f^{0}$ taken in $\mathbf{u}_>^0$ direction). 
	The curvature in $\mathbf{u}_>^{eq}$ direction is shown 
	by the dashed lines.
	Dots give the position of critical points. 
	%The thick curve indicates fluctuations in the liquid for the case of equilibrium 
	%(curvature taken in $\mathbf{u}_{coL}$ direction at the liquid point). 
 	} 
\label{fluc-3F}
\end{figure}

As we can see from fig.\ref{fluc-3F}, this 
%
%cor061009
%fluctuation 
inverse curvature
%cor061009
%
is not negligible 
but rather comparable to the case of equilibrium.
In this figure, we compare the cases of equilibrium and spinodal decomposition for the three different Skyrme parameterizations. We consider systems along the axis joining the critical points. The abscissa is the isovector density of the global system, $\rho_3=\rho_n-\rho_p$. 
For each point, we represent the $\rho_3$ projection of 
the different inverse curvatures. 
For spinodal decomposition, the relevant free energy 
is $f^{0}$, and its inverse curvature is calculated in $\mathbf{u}_>^{0}$ direction. At equilibrium, the concave intruder in $f^{0}$ is eliminated by Gibbs construction, 
and the inverse curvature of the resulting free energy $f^{eq}$ is computed 
%PC at the corresponding liquid point in $\mathbf{u}_{coL}$ direction. 
%PC The inverse curvature of $f^{eq}$ is also computed 
in $\mathbf{u}_>^{eq}$ direction for the global system,
which can be directly compared to the convexity of the unstable 
homogeneous system.

The main features obtained do not depend on Skyrme parameterization. 
We can see that, inside spinodal region, the convexification of the free
energy does not modify drastically the curvature in the isospin direction.
If spinodal decomposition is the 
dominant mechanism of phase separation, the isospin fluctuation tends to decrease
with increasing global asymmetry, as in the equilibrium case.
A complete dynamical treatment of the problem is needed to have quantitative 
predictions for the final fragment isotopic widths \cite{Liu,ColonnaMatera}. 
However in the minimal hypothesis
of a completely diabatic dynamics that propagates the initial fluctuation over 
the free energy surface, we can expect 
%PC a similar amplification factor of the 
%PC initial fluctuation as in the equilibrium case. 
%PC As a consequence, 
the absolute isotopic width of fragments issuing from
spinodal decomposition %PC is expected 
to be 
comparable
%, or even higher than in the case of coexisting phases at equilibrium.
to the case of coexisting phases at equilibrium.
The high asymmetry region where this trend is inversed is not physical, since
the gaussian approximation breaks down there for the liquid fraction 
in the equilibrium case.
 %It shows that both phase equilibrium and spinodal decomposition lead to fluctuations in the phase composition, with the same order of magnitude. Considering the low-asymmetry region, we see that the isospin fluctuations in spinodal decomposition take values between isospin fluctuations in the global equilibrated system and in the liquid part, where they are more important.

%%%%%%%%%%%%%%%%%%%%%%%%%%%%%%%%%%%%%%%%%%%%%%%%%%%%%%%%%%%%%%%%%%%%%
%%%%%%%%%%%%%%%%%%%%%%%%%%%%%%%%%%%%%%%%%%%%%%%%%%%%%%%%%%%%%%%%%%%%%
%%%%%%%%%%%%%%%%%%%%%%%%%%%%%%%%%%%%%%%%%%%%%%%%%%%%%%%%%%%%%%%%%%%%%
%
%cor061009
To get more insight on this issue,
we can compare the (initial) 
fluctuations expected for the spinodal mechanism with the predictions of the equilibrium SMM model in the grancanonical approximation.
Within this formalism the production probability of an isotope composed of Z protons and N neutrons reads
\begin{equation}
P_{\beta,\mu_n,\mu_p}
\propto
e^ {\beta \left (-F_{\beta}(N,Z)+\mu_n N +\mu_p Z \right)}
\label{EQ:distrib-NZ}
\end{equation}

Let us denote $A_3=V\rho_3=N-Z$.
The isospin variance of the distribution (\ref{EQ:distrib-NZ}) is
$\left(\sigma_{A_3}^{eq}\right)^2=<A_3^2>-<A_3>^2=V^2\left(\sigma_{\rho_3}^{eq}\right)^2$.
Using the same saddle point approximation we used to get eq.(\ref{fluct}),
this is related to the free-energy curvature by
\begin{equation}
\left(\sigma_{A_3}^{eq}\right)^2
=\frac{1}{\beta}\left( \frac{\partial^2F}{\partial A_3^2} \right)^{-1}
=\frac{A}{2 \beta C_{sym}(A)}
\label{EQ:Sigma_A3_eq}
\end{equation}
%
%
%cor061108
%where $C_{sym}(A)\approx C_{sym}(\rho_0)$ is the symmetry energy of the produced fragment.
where $C_{sym}(A)$ is the symmetry energy of the produced fragment.
Denoting $\rho^{eq}$ the freeze-out fragment density,
% (with no surface effect)?
we have $C_{sym}(A)=C_{sym}(\rho^{eq})$.
%cor061108
%

In the minimal hypothesis that the initial fluctuations $\left(\sigma_{A_3}^{sd}\right)_i^2$
are not amplified (per unit volume) by the successive dynamics of phase separation :
\begin{equation}
\left(\sigma_{A_3}^{sd}\right)_f^2/V_f
=\left(\sigma_{A_3}^{sd}\right)_i^2/V_i
=V_i\left(\sigma_{\rho_3}^{sd}\right)_i^2 ,
\end{equation}
the final variance in the case of spinodal decomposition $\left(\sigma_{A_3}^{sd}\right)_f^2$
reads (\emph{cf} eq.(\ref{fluct}) and (\ref{fluct-Csym})) :
\begin{equation}
\left(\sigma_{A_3}^{sd}\right)_f^2
%
%cor061108
=\frac{V_f \rho_i}{2 \beta C_{sym}(\rho_i)}
%\approx \frac{A}{2\beta C_{sym}(\rho)}\frac{\rho}{\rho_0}
=\frac{A}{2\beta C_{sym}(\rho_i)}\frac{\rho_i}{\rho_f^{sd}}
\label{EQ:Sigma_A3_sd}
\end{equation}
where $\rho_i$ is the density at the onset of fragment formation
and $\rho_f^{sd}$ the freeze-out fragment density in a spinodal decomposition scenario.
Equations (\ref{EQ:Sigma_A3_eq}) and (\ref{EQ:Sigma_A3_sd}) lead to the relation :
\begin{equation}
\label{EQ:rapport-fluc}
\frac{\left( \sigma_{A_3}^{sd} \right)^2_f }{\left( \sigma_{A_3}^{eq} \right)^2}
=
\frac{C_{sym}(\rho^{eq})}{C_{sym}(\rho_i)}\frac{\rho_i}{\rho_f^{sd}}
\end{equation}
%
%cor061108
%
Using the standard representation \cite{baoanli}
$C_{sym}(\rho)=C_{sym}(\rho_0)(\rho/\rho_0)^{\gamma}$
%
%cor061108
and considering $\rho_f^{sd} \approx \rho^{eq} \approx \rho_0$,
%cor061108
%
we get
\begin{equation}
\label{EQ:rapport-fluc-2}
\frac{\left( \sigma_{A_3}^{sd} \right)^2_f }{\left( \sigma_{A_3}^{eq} \right)^2}
\approx \left( \frac{\rho}{\rho_0}\right)^{1-\gamma}
\end{equation}
which can be even greater than $1$ for an asy-stiff equation of state.

Considering that in the dynamics of spinodal decomposition isovector density fluctuations
will also be (slightly) increased from the initial condition 
%eq.(\ref{fluct}) 
\cite{ColonnaMatera}, eq.(\ref{EQ:rapport-fluc-2})
confirms that the fluctuations associated with the two scenarios are expected to be comparable.
%cor061009
%

%%%%%%%%%%%%%%%%%%%%%%%%%%%%%%%%%%%%%%%%%%%%%%%%
%%%%%%%%%%%%%%%%%%%%%%%%%%%%%%%%%%%%%%%%%%%%%%%%
%%%%%%%%%%%%%%%%%%%%%%%%%%%%%%%%%%%%%%%%%%%%%%%%

\section{Conclusion}

In this paper, we have presented isospin properties of the phases formed in nuclear-matter liquid-gas transition. The characteristics of phase separation are deduced from the free-energy curvature properties, $f$ corresponding to the constrained entropy at a fixed temperature $\beta$. Under a critical temperature, the free energy of the homogeneous system $f^{0}$ presents a region of abnormal curvature, where phase separation is favorable. Phase equilibrium is obtained according to Gibbs construction, determining an equilibrated free energy $f^{eq}$. Phase properties are then deduced from the free-energy curvature matrix, studying $f^{eq}$ for a system at equilibrium and $f^{0}$ for a system undergoing spinodal decomposition. The direction of phase separation is given by the eigen-vector $\mathbf{u}_<$ linked to the lower eigen-value
of the curvature matrix. %PC <<<<
 In both equilibrium and spinodal decomposition, it leads to isospin fractionation, with a liquid more symmetric than the global system. In the direction orthogonal to phase separation, the curvature is linked to fluctuations affecting phase composition. This direction being close to $\rho_3$ direction, it is strongly linked to isospin fluctuations. 
Spinodal decomposition is predicted to give a stronger average fractionation and 
at least comparable or even higher isospin fluctuations with respect to a phase separation at equilibrium.
Isoscaling observables as well as isotopic distributions are expected to be 
sensitive to such effects. If fragmentation can be associated with a spinodal decomposition,
we 
%
%cor061009
qualitatively
%cor061009
%
expect i) low apparent values of the symmetry energy coefficient extracted 
from isoscaling analyses, and ii) isotopic widths comparable or even larger than in 
the case of the statistical model.
%
%cor061009
%However, finite-size corrections have to be introduced before
%quantitative predictions can be done.
Finally we would like to stress that the present work concerns an idealized study  of bulk matter, which  is likely subject to significant modification when the nuclear surface is taken into account. 
The pursuit of this issue would be a very interesting topic for further study.
%cor061009
%
%It appears that such fluctuations are the same order of magnitude at equilibrium and in spinodal decomposition.

%%%%%%%%%%%%%%%%%%%%%%%%%%%%%%%%%%%%%%%%%%%%%%%%
%%%%%%%%%%%%%%%%%%%%%%%%%%%%%%%%%%%%%%%%%%%%%%%%
%%%%%%%%%%%%%%%%%%%%%%%%%%%%%%%%%%%%%%%%%%%%%%%%
% BIBLIO

\end{document}